\documentclass[prb,twocolumn,showpacs,a4paper,superscriptaddress]{revtex4}

\usepackage{amsmath}
\usepackage{amsfonts}
\usepackage{amssymb}
\usepackage{gensymb}
\usepackage{textcomp}
\usepackage{psfrag,graphicx}
\usepackage{subfigure}
\usepackage{picture,xcolor}

\providecommand*{\unit}[1]{\,\ifmmode \mathrm{\,#1}\else\textup{#1}\fi}

\begin{document}

\title{Leakage Current of a Superconductor-Normal Metal Tunnel Junction Connected to a High-Temperature Environment}

\author{A. Di Marco}
\affiliation{LPMMC-CNRS, Universit\'{e} Joseph Fourier, 25 Avenue des Martyrs BP166 38042 Grenoble Cedex, France}

\author{V. F. Maisi} \affiliation{Centre for Metrology and Accreditation (MIKES),
P.O. Box 9, 02151 Espoo, Finland} \affiliation{Low Temperature Laboratory, Aalto University, P.O.
Box 13500, FI-00076 Aalto, Finland}

\author{J. P. Pekola}
\affiliation{Low Temperature Laboratory, Aalto University, P.O. Box 13500, FI-00076 Aalto, Finland}

\author{F. W. J. Hekking}
\affiliation{LPMMC-CNRS, Universit\'{e} Joseph Fourier, 25 Avenue des Martyrs BP166 38042 Grenoble Cedex, France}

\date{\today}

\begin{abstract}
We consider a voltage-biased Normal metal-Insulator-Superconductor (NIS) tunnel junction,
connected to a high-temperature external electromagnetic environment. This model system
features the commonly observed subgap leakage current in NIS junctions through
photon-assisted tunneling which is detrimental for applications. We first consider a NIS
junction directly coupled to the environment and analyze the subgap leakage current both
analytically and numerically; we discuss the link with the phenomenological Dynes
parameter. Then we focus on a circuit where a low-temperature lossy transmission line is
inserted between the NIS junction and the environment. We show that the subgap leakage
current is exponentially suppressed as the length, $\ell$, and the resistance per unit
length, $R_0$, of the line are increased. We finally discuss our results in view of the
performance of NIS junctions in applications.
\end{abstract}

\pacs{74.55.+v,74.25.F-,85.25.Am,72.70.+m}

\maketitle

\section{Introduction}

The peculiar nature of single-particle electronic transport through a normal metal-insulator-superconductor (NIS)
junction is at the origin of several interesting applications. Such junctions are widely used in experiments of
mesoscopic physics as a spectroscopic tool\cite{spectrum1,spectrum2}, as a very sensitive thermometer
\cite{cooling,thermo1,thermo2} and as a key element in nano-refrigeration \cite{cooling,cooling1,cooling2}.
Furthermore, NIS junctions are currently investigated in view of achieving a high accuracy when controlling the current
through a single-electron SINIS turnstile. Such a device is one of the interesting candidates for the completion of the
so-called quantum metrological triangle, \emph{i.e.}, it can be used to obtain a precise realization of
current\cite{SINIS1,SINIS2}. These applications are all based on the existence of the Bardeen-Cooper-Schrieffer (BCS)
energy gap $\Delta$ in the density of states (DoS) of the superconductor\cite{BCS}. Ideally one would expect no
single-electron current to flow through a NIS junction at low temperature as long as the bias voltage $V$ satisfies the
inequality $-\Delta<eV<\Delta$.

In practice the subgap current is different from zero. This is a central problem which
limits the performance of applications based on energy-selective single-particle
transport in NIS junctions. The presence of unwanted accessible states in the subgap
region manifests itself as a smearing of the junction's current-voltage ($I$-$V$)
characteristic as well as of its differential conductance. Giaever was the first to
experimentally study the NIS junction. He noticed that this deviation from the ideal
behavior was present even if the junction was kept at a temperature much lower than the
critical one $T_c$ of the superconductor\cite{Giaever}. A possible source of subgap
leakage currents is the occurrence of many-electron tunneling processes, such as Andreev
reflection \cite{andreev1,andreev2,andreev3}. However, these many-electron processes are
strongly suppressed if the tunnel resistance $R_T$ of the junction is chosen high enough
and do not account for the observed residual subgap transport either.

\begin{figure*}[ht!]
\centering \subfigure[]
   {\begin{psfrags}\psfrag{V}[tc][tc][1.5]{$V$}
   \psfrag{C}[tc][tc][1.5]{$C$}
   \psfrag{Rt}[tc][tc][1.5]{$R_T$}
   \psfrag{Tj}[tc][tc][1.5]{$T_\textrm{jun}$}
   \psfrag{Te}[tc][tc][1.5]{$T_\textrm{env}$}
   \psfrag{Z}[tc][tc][1.5]{$Z(\omega)$}
   \includegraphics[height=3.6cm]{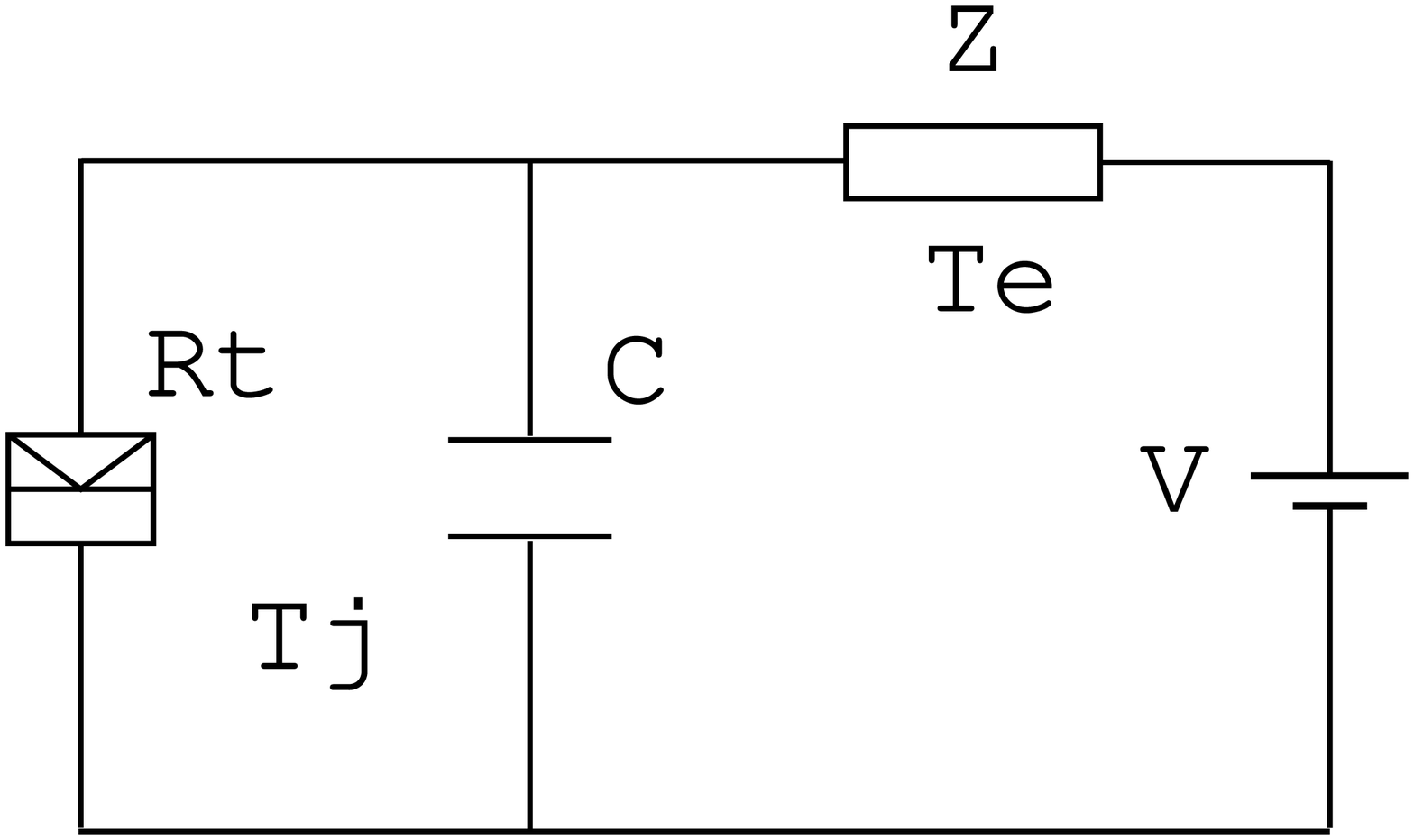}\label{fig1}\end{psfrags}}
\hspace{7mm} \subfigure[]
   {\begin{psfrags}\psfrag{V}[tc][tc][1.5]{$V$}
   \psfrag{C}[tc][tc][1.5]{$C$}
   \psfrag{Rt}[tc][tc][1.5]{$R_T$}
   \psfrag{Tj}[tc][tc][1.5]{$T_\textrm{jun}$}
   \psfrag{Tt}[tc][tc][1.5]{$T_\textrm{line}$}
   \psfrag{Te}[tc][tc][1.5]{$T_\textrm{env}$}
   \psfrag{Z}[tc][tc][1.5]{$Z(\omega)$}
   \psfrag{l}[tc][tc][1.5]{$\ell$}
   \psfrag{par}[tc][tc][1.5]{$R_0\, , C_0\, , L_0$}
   \includegraphics[height=3.6cm]{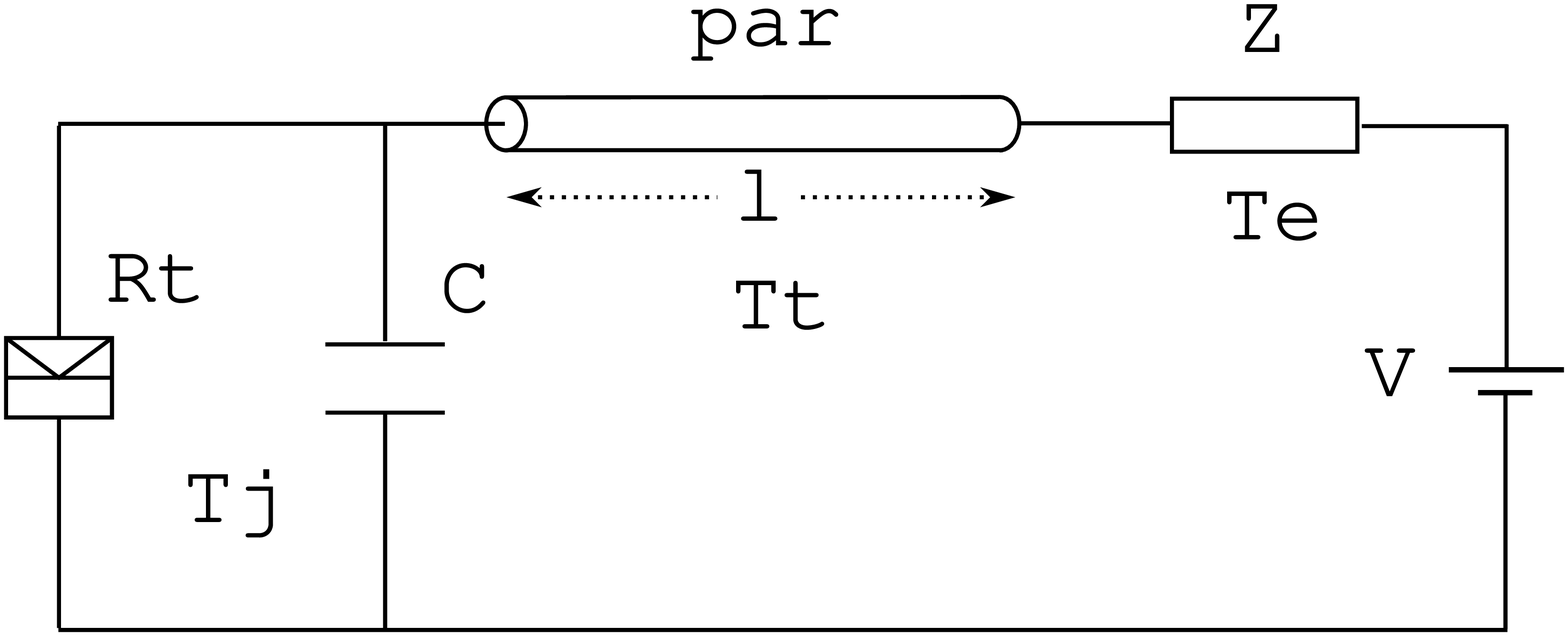}\label{fig2}\end{psfrags}}
\caption{Circuit representation of the two configurations studied in this paper. (a) A
NIS junction at temperature $T_\textrm{jun}$ is connected in parallel to its capacitance
$C$ and to an impedance $Z(\omega)$ which represents the high-temperature environment at
temperature $T_\textrm{env}\gg T_\textrm{jun}$. The whole circuit is biased by the
constant voltage $V$. (b) A transmission line of length $\ell$ is inserted between the
junction and the impedance $Z(\omega)$ of circuit (a). It is described by the parameters
$R_0$, $C_0$ and $L_0$, the resistance, the capacitance and the inductance per unit
length, respectively, as well as by its temperature which is assumed equal to
$T_\textrm{jun}$.}
\end{figure*}

Dynes modified the BCS superconducting DoS introducing a phenomenological dimensionless parameter,
$\gamma_\textrm{\tiny{Dynes}}$, in order to fit the behavior of the subgap quasi-particle tunneling current through a
Josephson junction\cite{Dynes}. The modified DoS, normalized to the corresponding normal-state DoS at the Fermi energy,
is given by
\begin{equation}
\label{DoS.Dynes}
N_S^\textrm{\tiny{Dynes}}=\Bigg|\Re\mbox{e}\Bigg[\frac{E/\Delta+i\gamma_\textrm{\tiny{Dynes}}}{\sqrt{(E/\Delta+
i\gamma_\textrm{\tiny{Dynes}})^2-1}}\Bigg]\Bigg|.
\end{equation}
It can be seen that $\gamma_\textrm{\tiny{Dynes}}$ indeed accounts for the broadening of the DoS around $\Delta$ and
the occurrence of states within the gap. This expression is frequently used in both numerical and analytical
calculations, but concerning the microscopic origin of the Dynes parameter $\gamma_\textrm{\tiny{Dynes}}$, for
temperatures far below $T_c$, relatively little is known.

Recently it was realized that the exchange of energy between the NIS junction and its surrounding electromagnetic
environment may give rise to subgap features similar to those accounted for phenomenologically by the Dynes parameter
\cite{main,environment1}. Indeed, under certain conditions, energy absorption from such an environment enables the
crossing of the tunnel barrier by single electrons even for $|V|$ much less than $\Delta/e$. Within this framework an
analytical expression for $\gamma_\textrm{\tiny{Dynes}}$ has been obtained in terms of parameters characterizing the
NIS junction's environment for the special case of a purely resistive external circuit\cite{main}.

Following the idea of photon-assisted tunneling demonstrated in Ref.~\onlinecite{main}, we generalize the approach here
and obtain expressions for the subgap leakage current and the Dynes parameter valid for an external circuit
characterized by an arbitrary impedance $Z(\omega)$, kept at a temperature $T_\textrm{env}$ that is not necessarily the
temperature $T_\textrm{jun}$ of the NIS junction, see Fig.~\ref{fig1}. Then we turn our attention to the circuit
depicted in Fig.~\ref{fig2}, where we study the effects of the insertion of a lossy transmission line, meant to act as
a filter, between the cold junction and the high-temperature external impedance $Z(\omega)$. In particular we use our
results to understand under which conditions the transmission line will behave as a filter capable of reducing the
photon-assisted tunneling induced by the high-temperature external impedance and thus reducing the Dynes parameter to
values that are compatible with the accuracy requirements for applications such as the SINIS turnstile.

\section{NIS Junction coupled to a high-temperature environment}\label{Sec.NIS}

\subsection{Single-Particle Current}

We start by considering the basic circuit illustrated in Fig.~\ref{fig1} where a NIS
junction is connected in series to an effective high-temperature impedance $Z(\omega)$.
The junction itself is characterized by a tunnel resistance $R_T$ in parallel with a
capacitance $C$. The entire circuit is voltage biased. This constitutes a minimal model
for a junction embedded in an external electromagnetic environment at temperature
$T_\textrm{env}$, which can be much higher than the temperature $T_\textrm{jun}$ of the
junction.

According to the so-called $P(E)$ theory\cite{IngNaz}, the single-particle tunneling current through a NIS junction
coupled to an external environment is given by
$$
I_{\textrm{NS}}(V)=\frac{1}{eR_T}\int dE \int dE' \ N_S\big( E' \big)\Big[1-f\big(E'\big)\Big]\times
$$
\begin{equation}
\label{NS.PE.current1}
\times \Bigg\{f\big(E-eV\big)-f\big(E+eV\big)  \Bigg\} \ P\big(E-E'\big) \ .
\end{equation}
Here, the energy $E$ refers to the electrons of the normal metal, $E'$ is the energy of the superconductor
quasi-particles, $N_S(E')$ is the BCS density of states of the superconducting wire divided by the normal-metal DoS at
the Fermi level and $f(E) = [e^{\beta_\textrm{jun} E} +1]^{-1}$ is the Fermi-Dirac distribution with
$\beta_\textrm{jun} = 1/k_B T_\textrm{jun}$ the inverse temperature of the junction. Expression (\ref{NS.PE.current1})
does not take into account the higher order processes in tunneling which will be ignored throughout this paper. The
validity of this assumption will be discussed in Sec.\ref{Sec.multi}.

The function $P(E)$ in Eq.~(\ref{NS.PE.current1}) is the probability density that the
tunneling electron exchanges an amount of energy $E$ with the environment. This process
takes place through the emission or absorption of photons. It is defined as
\begin{equation}\label{PE}P(E)=\frac{1}{2\pi \hbar}\int_{-\infty}^{+\infty}dt \ e^{iEt/\hbar} \
e^{J(t)} \ ,
\end{equation}
\emph{i.e.}, it is the Fourier transform of the exponential of the correlation function
$$
J(t)=2\int_0^{+\infty}\frac{d\omega}{\omega} \ \frac{\Re\mbox{e}\big[ Z_\textrm{tot}(\omega)\big]}{R_K}\times
$$
\begin{equation}\label{J}\times\Bigg\{ \coth\Big(
\frac{1}{2}\beta_{\textrm{env}}\hbar\omega\Big)\Big[ \cos\big( \omega t \big) -1   \Big] -i\sin\big( \omega t \big)
\Bigg\} \ .
\end{equation}
Here $Z_\textrm{tot}(\omega)$ is the total impedance seen by the junction, resulting from the connection in parallel of
$C$ and $Z(\omega)$, $R_K=h/e^2$ is the quantum resistance and $\beta_\textrm{env}=1/k_BT_\textrm{env}$.

The function $J(t)$ determines the strength of the coupling between the NIS junction and the environment. Indeed if
$J(t)=0$, the probability density $P(E)$ is equal to a Dirac delta $\delta(E)$ and the single-particle tunneling
current is elastic. Expression (\ref{NS.PE.current1}) then reduces to the standard expression for single-particle
tunneling in NIS junctions valid in the absence of environment. The environment-induced inelastic tunneling processes
occur only when $J(t)\neq0$. In general, the time intervals where the inelastic effects are important are related to
the energy ranges where $P(E)\neq 0$. Depending on the value of $J(t)$, the coupling between the NIS junction and the
environment can be considered weak or strong. Throughout this paper we will treat both regimes of weak and strong
coupling in more detail.

In order to analyze the smearing of the NIS junction's $I$-$V$ characteristic due to the presence of the
high-temperature environment, we will ignore the thermal smearing induced by finite temperature of the N and S
electrodes. This is an adequate approximation under standard experimental conditions where $T_\mathrm{jun} \ll
\Delta/k_B$. Hereafter we will set the temperature of the junction $T_\textrm{jun}$ to zero. Under this assumption the
single-particle current (\ref{NS.PE.current1}) becomes
\begin{equation}
\label{NS.PE.current2}
I_{\textrm{NS}}(V)\simeq\frac{1}{eR_T}\int_{-eV}^{+eV}
dE \int_{\Delta}^{+\infty} dE' \ N_S\big( E' \big) \ P\big(E-E'\big) \ .
\end{equation}
We furthermore will focus on the subgap region of the $I$-$V$ curve considering $|eV| \ll \Delta$. As a result, the
integration variables $|E| \ll E'$ in (\ref{NS.PE.current2}), and we can approximate $P(E-E') \approx P(-E')$. The
resulting integral over $E$ can be performed immediately to yield
\begin{equation}
\label{NS.PE.current3} I_{\textrm{NS}}^{\textrm{sub}}(V)\simeq \ \Gamma \ \frac{V}{R_T} \ ,
\end{equation}
where the factor $\Gamma$ is given by the integral
\begin{equation}
\label{gamma}
\Gamma=2\int_{\Delta}^{+\infty} dE' \ N_S\big( E' \big) \ P\big(-E'\big) \ .
\end{equation}
We see that for the parameter $\Gamma$, Eq.~(\ref{gamma}), and hence the subgap current given by
Eq.~(\ref{NS.PE.current3}) to be nonzero, the function $P(E)$ should be nonzero for energies $E \le -\Delta$. This
reflects the fact that under subgap conditions $eV, k_BT_\textrm{jun} \ll \Delta$,  a nonzero single-particle current
occurs only if the tunneling electrons absorb an energy $\agt \Delta$ from the environment. For instance, $\Gamma = 0$
for elastic tunneling in the absence of an environment, when $P(E)= \delta(E)$. We also expect $\Gamma$ to vanish when
the temperature of the environment $k_BT_\textrm{env}$ is much less than the energy gap $\Delta$. Indeed, due to
detailed balance\cite{IngNaz}, $P(-E)=e^{-E/k_BT_\textrm{env}}P(E)$, the function $P(E)$ is strongly suppressed for
negative energies $E < - k_BT_\textrm{env}$. This means that the integral in (\ref{gamma}) will vanish unless the
environment is sufficiently hot, $k_BT_\textrm{env}\gtrsim \Delta$.

In order to make a connection with the aforementioned approach due to Dynes, we linearize the usual expression for
elastic single-particle tunneling in a NIS junction, using the Dynes DoS (\ref{DoS.Dynes}) to characterize the
superconducting electrode. One obtains the linear subgap current-voltage relationship
$$
I_{\textrm{NS}}^{\textrm{sub}}(V)\simeq \
\sqrt{\frac{\gamma_\textrm{\tiny{Dynes}}^2}{\gamma_\textrm{\tiny{Dynes}}^2+1}} \ \frac{V}{R_T} \ .
$$
Comparing this result with Eq.~(\ref{NS.PE.current3}) above, we conclude that, in the
linear regime, the Dynes parameter can be related to $\Gamma$, according to $\Gamma =
\sqrt{\gamma_\textrm{\tiny{Dynes}}^2/(\gamma_\textrm{\tiny{Dynes}}^2+1)}$. We see in
particular that the two parameters coincide $\gamma_\textrm{\tiny{Dynes}} = \Gamma$
whenever $\Gamma, \gamma_\textrm{\tiny{Dynes}}\ll1$. This shows that fluctuations of a
high-temperature electromagnetic environment constitute a possible microscopic source of
the phenomenological Dynes parameter.

\begin{figure*}[!ht]
\newcommand{\myfbox}{\fbox}
\centering \subfigure[] {\begin{psfrags} \psfrag{Strong}[tc][tc][1.1]{Strong} \psfrag{Weak}[tc][tc][1.1]{Weak}
\psfrag{temp}[tc][tc][1]{$\frac{\hbar/k_BT_\textrm{env}}{R_K C}$} \psfrag{delta}[tc][tc][1]{$\frac{\hbar/\Delta}{R_K
C}$} \psfrag{tau}[tc][tc][1.7]{$\tau_\textrm{\tiny{10\%}}(\rho)$}
\psfrag{rho}[tc][tc][1.7]{$\rho$}\psfrag{0}[tc][tc][1.5]{$0$}\psfrag{rhot}[tc][tc][1.5]{$\rho_\textrm{\tiny{th}}$}
\psfrag{rhod}[tc][tc][1.5]{$\rho_{\scriptscriptstyle\Delta}$}\psfrag{taus}[tc][tc][1.5]{$\tau_\textrm{\tiny{S}}$}
\psfrag{taue}[tc][tc][1.5]{$\tau_\textrm{e}$}\psfrag{taud}[tc][tc][1.5]{$\tau_{\scriptscriptstyle\Delta}$}
{\includegraphics[trim=30 45 15 3,height=0.28\textwidth]{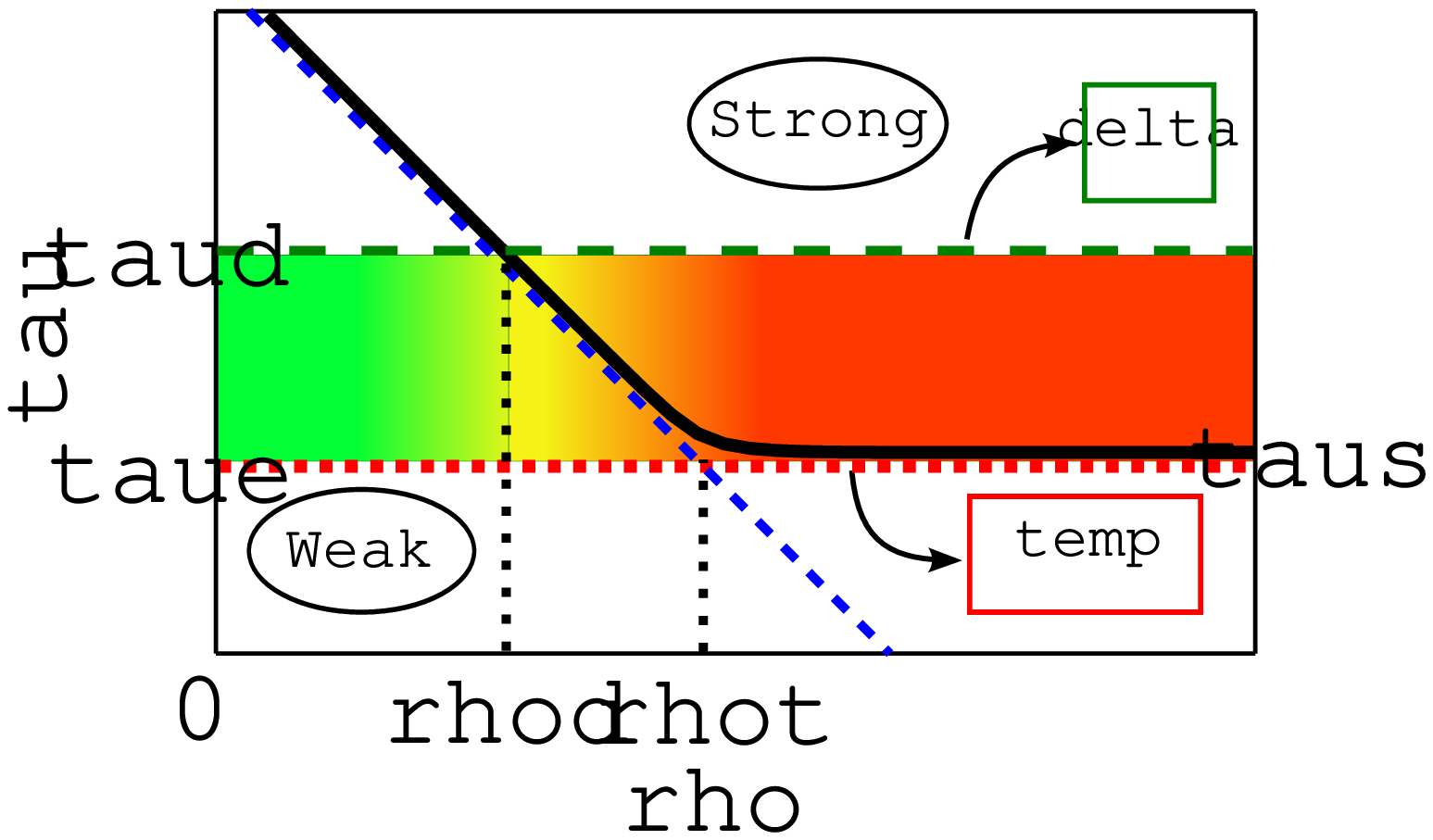}\label{fig:taurho2}}\end{psfrags}}\qquad
 \hspace{5mm}\subfigure[] {\begin{psfrags} \psfrag{01}[tc][tc][0.85]{$\quad\qquad 0.005 \unit{K}$}
\psfrag{02}[tc][tc][0.85]{$\quad\qquad 0.05 \unit{K}$} \psfrag{03}[tc][tc][0.85]{$\quad\qquad 0.5 \unit{K}$}
\psfrag{04}[tc][tc][0.85]{$\quad\qquad 5 \unit{K}$} \psfrag{05}[tc][tc][0.85]{$\quad\qquad 50 \unit{K}$}
\psfrag{Leglab}[tc][tc][1]{$T_\textrm{env}$}
\psfrag{xxx}[bc][bc][1.7]{$\rho$}\psfrag{yyy}[bc][bc][1.7]{$\tau_\textrm{\tiny{10\%}}(\rho)$} {\includegraphics[trim=5
6 30 7,height=0.28\textwidth]{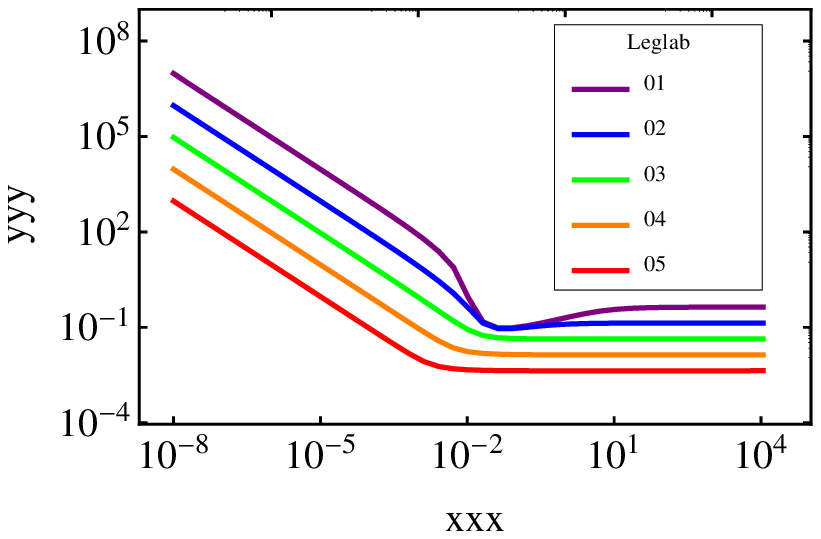}}\label{fig:taurho}
\end{psfrags}}
\caption{Plot of the separatrix $\tau_\textrm{\tiny{10\%}}(\rho)$ as a function of the dimensionless resistance
$\rho=R/R_K\,$, defined as the solution of the equation $\exp{\big\{\Re\mbox{e}\big[J(\tau_\textrm{\tiny{10\%}},\rho)
\big]}\big\}=0.9$ with $C$ fixed. Both plots are in double logarithmic scale. (a) For a fixed value of
$T_\textrm{env}\,$,  $\tau_\textrm{\tiny{10\%}}(\rho)$ separates the weak and strong coupling regions (black thick
line). The colored strip indicates the time interval bound by  $\tau_{\scriptscriptstyle\Delta}=\hbar/\Delta R_K C$
(dark green dashed line) and $\tau_\textrm{e}=\hbar/k_BT_\textrm{env}R_K C$ (red dashed line). The intersection between
$\tau_{\scriptscriptstyle\Delta}$ and the separatrix $\tau_\textrm{\tiny{10\%}}(\rho)$ defines the resistance
$\rho_\Delta\,$. The asymptotic expression for $\tau_\textrm{\tiny{10\%}}(\rho)$ valid for $\rho\rightarrow0$ and
proportional to $1/\rho$ is also shown (blue dashed line). Its intersection with the line corresponding to
$\tau_\textrm{\tiny{S}}\sim\sqrt{\hbar/k_BT_\textrm{env}R_K C}$ defines the threshold resistance $\rho_\textrm{th}$. On
the logarithmic scale used here, $\tau_\textrm{\tiny{S}}$ almost coincides with $\tau_\mathrm{e}$. (b) As the
temperature of the environment, $T_\textrm{env}\,$, is decreased, the curve $\tau_\textrm{\tiny{10\%}}(\rho)$ moves up,
thereby increasing the weak coupling region.}\label{fig:weakTauRho}
\end{figure*}

\subsection{Weak and Strong Coupling}\label{Sec.NIS.w&s}

As we have seen above, the strength of the coupling between the NIS junction and the environment is determined by the
function $J(t)$. Let us assume that this function is small, in a sense to be detailed below. Expanding the exponential
function $\exp[J(t)]$ up to the first order in $J(t)$, Eq.~(\ref{PE}) becomes
\begin{equation}
\label{PE.exp}P(E) \simeq\frac{1}{2\pi\hbar}\int_{-\infty}^{+\infty}dt \ e^{iEt/\hbar} \ \Big[ 1+J(t)\Big] \ .
\end{equation}
The evaluation of the integral over time in (\ref{PE.exp}) gives
$$
P(E)\simeq \ \delta\big(E\big)+\frac{1}{\hbar}\int_0^{+\infty}\frac{d\omega}{\omega} \ \frac{\Re\mbox{e}\big[
Z_{\textrm{tot}}(\omega)\big]}{R_K} \ \times
$$
$$
\times\Bigg\{
 \ \Bigg[\coth\Big(\frac{1}{2}\beta_{\textrm{env}}\hbar\omega\Big)-1\Bigg] \
\delta\Big(\frac{E}{\hbar}+\omega\Big)+
$$
$$
+\Bigg[\coth\Big(\frac{1}{2}\beta_{\textrm{env}}\hbar\omega\Big)+1\Bigg] \ \delta\Big(\frac{E}{\hbar}-\omega\Big)-
$$
\begin{equation}
\label{PEweak.one} -2\hbar \ \coth\Big(\frac{1}{2}\beta_{\textrm{env}}\hbar\omega\Big) \ \delta\big(E\big) \ \Bigg\} \
.
\end{equation}
We see that the function $P(E)$ has an elastic contribution and an inelastic one involving the exchange of at most one
photon between the junction and the environment. In fact the first and the fourth terms represent the elastic tunneling
involving zero and one virtual photon, respectively. The second and third terms are related to the process of
absorbtion and emission of one real photon, respectively. We define this one-photon regime as weak coupling. On the
other hand, the coupling becomes strong whenever the single-photon exchange between the junction and the environment is
no longer the dominant effect. In this case, the higher-order terms cannot be neglected in the series expansion of
$\exp\big[J(t) \big]\,$, indicating that multi-photon processes have to be taken into account.

We proceed by determining the time interval where the expansion (\ref{PE.exp}) holds. Given the fact that $J(t=0) = 0
$, we expect this to be the short time interval~\cite{IngNaz}. We set $Z(\omega)=R$ for simplicity and introduce the
dimensionless time $\tau=t/R_KC$ as well as the ratio $\rho=R/R_K$. The quantity
$\exp{\big\{\Re\mbox{e}\big[J(\tau,\rho) \big]}\big\}$ decays monotonically with increasing time $\tau$, starting from
unity at $\tau = 0$. The rate at which it decays depends on $\rho$: the larger $\rho$, the faster it decays, in
agreement with Ref.~\onlinecite{IngNaz}. We determine the relevant short time interval by determining the
characteristic time $\tau_\textrm{\tiny{10\%}}$, at which the quantity $\exp{\big\{\Re\mbox{e}\big[J(\tau,\rho)
\big]}\big\}$ dropped by $10 \%$. Figure \ref{fig:taurho2} shows $\tau_\textrm{\tiny{10\%}}$ as a function of the
parameter $\rho$, keeping $T_\textrm{env}$ and $C$ fixed. The line $\tau_\textrm{\tiny{10\%}}(\rho)$ separates the weak
coupling regime found at short times from the strong coupling regimes reached for longer times. As
expected~\cite{IngNaz}, with increasing $\rho$, the separatrix $\tau_\textrm{\tiny{10\%}}(\rho)$ decreases as $1/\rho$,
and then saturates at a value $\tau_\textrm{\tiny{S}} \sim \sqrt{\hbar/k_B T_\mathrm{env} R_K C}$ for $\rho >
\rho_\textrm{th} \sim \tau_\textrm{\tiny{S}}$. As shown in Fig.~\ref{fig:taurho}, the curve
$\tau_\textrm{\tiny{10\%}}(\rho)$ shifts up when decreasing the temperature of the environment, $T_\textrm{env}\,$,
thereby increasing the time interval where the expansion (\ref{PE.exp}) holds.

We now return to the inelastic tunneling of single electrons through the NIS junction. Under subgap conditions $k_B
T_\textrm{jun} ,  eV \ll \Delta\,$, the energy $E$ relevant for the photon-assisted tunneling processes is in the
interval $\Delta \lesssim  E \lesssim k_BT_\textrm{env}\,$. The upper bound corresponds to the largest energy the
junction can absorb from the environment. In time domain, we thus have to consider the interval $\tau_\mathrm{e} < \tau
< \tau_\Delta$ where $\tau_\Delta=\hbar/\Delta R_K C$ and $\tau_\mathrm{e}= \hbar/k_BT_\textrm{env}R_K C$. This
interval is represented by the colored strip in Fig.~\ref{fig:taurho2}. Note that on the logarithmic scale used here,
the lower bound $\tau_\mathrm{e}$ almost coincides with the value $\tau_\mathrm{S}$ at which the separatrix saturates
for large values of $\rho$. The intersection between $\tau_\Delta$ and the $10\%$ curve
$\tau_\textrm{\tiny{10\%}}(\rho)\,$ defines the characteristic resistance $\rho_\Delta\,$ separating the weak and
strong coupling regimes. When $\rho<\rho_\Delta$, coupling is weak and only single-photon absorption processes occur
(green area); if $\rho\sim \rho_\Delta$ both single- and multi-photon processes occur during single-electron tunneling
(yellow-orange area); as soon as $\rho\gg\rho_\Delta$, multi-photon processes become dominant (red area). In
particular, the two limiting cases $\rho\ll\rho_\Delta\, , \rho_\textrm{th}\,$ and $\rho\gg\rho_\Delta\, ,
\rho_\textrm{th}\,$ are equivalent to the conditions $R/R_K\ll \Delta/k_BT_\textrm{env}$ and $R/R_K\gg
\Delta/k_BT_\textrm{env}$ respectively.

\subsection{Subgap Leakage Current: Weak Coupling}\label{Sec.NIS.weak}

We start by dealing with the weak coupling case. Since we are interested in the subgap
region of the $I$-$V$ characteristic, $k_BT_\mathrm{jun}, eV\ll\Delta$, the behavior of
the function $P(E)$ at energies $E>-\Delta$ is irrelevant. Therefore we can ignore the
elastic contributions in Eq.~(\ref{PEweak.one}). Evaluating the integral over frequencies
in Eq.~(\ref{PEweak.one}), the relevant contribution to the function $P(E)$ for energies
$E \ne 0$ reads
\begin{equation}
\label{NS.PE.weak}
P(E)\simeq \  2 \
\frac{\Re\mbox{e}\big[Z_\textrm{tot}\big(E/\hbar\big)\big]}{R_K}\Bigg(\frac{1+n\big(E\big)}{E} \Bigg) \ .
\end{equation}
Here $n(E)=[e^{\beta_\textrm{env}E}-1]^{-1}$ is the Bose-Einstein distribution of the
photons of the environment.

\begin{figure}[t]
\newcommand{\myfbox}{\fbox}
\psfrag{01}[tc][tc][0.85]{$\quad 0.1$} \psfrag{02}[tc][tc][0.85]{$\quad 0.5$} \psfrag{03}[tc][tc][0.85]{$\quad 1$}
\psfrag{04}[tc][tc][0.85]{$\quad 5$} \psfrag{Leglab}[tc][tc][0.9]{$\Delta RC/\hbar$}
\psfrag{xxx}[bc][bc][1.3]{$k_BT_\textrm{env}/\Delta$}\psfrag{yyy}[Bc][Bc][1.3]{$R_K \Gamma/R$}
\includegraphics[trim=0 5 30 4,height=0.31\textwidth]{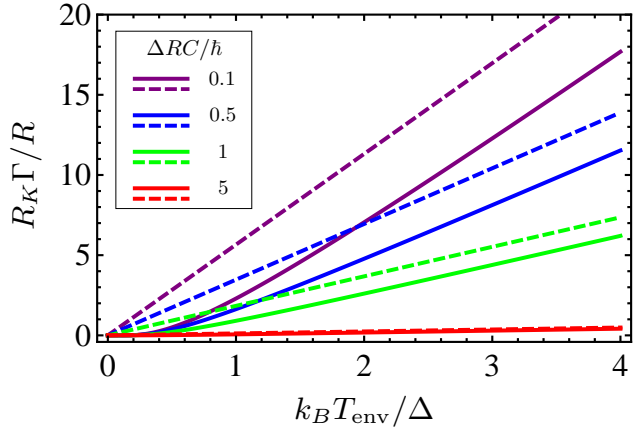} \caption{Plot of the rescaled parameter
$R_K \Gamma/R$ as a function of $k_B T_\mathrm{env}/\Delta$ for different values of the
ratio $\Delta RC/\hbar$. Solid lines are obtained by a numerical integration of
Eq.~(\ref{gamma.weak}) using Eq.~(\ref{totresR}). Dashed lines refer to the asymptotic
$\Gamma$ given by Eq.~(\ref{gamma.weak.a}).}\label{fig:resistives}
\end{figure}

The probability density (\ref{NS.PE.weak}) can be used to get a limiting expression for
$\Gamma$,
\begin{equation}
\label{gamma.weak} \Gamma= \ 4\int_\Delta^{+\infty} dE \ N_S(E) \ \frac{\Re\mbox{e}\big[
Z_{\textrm{tot}}\big(E/\hbar\big)\big]}{R_K} \ \frac{n\big(E\big)}{E} \ .
\end{equation}
Let us apply this result to the example of a purely resistive external environment. This model has been used before to
study devices based on tunnel junctions in connection with experiments\cite{main,purelyR1,purelyR2}. Replacing the
external impedance $Z(\omega)$ of the circuit of Fig.~\ref{fig1} by a pure resistance $R$, the real part of the total
impedance is
\begin{equation}
\label{totresR}
\Re\mbox{e}\big[
Z_{\textrm{tot}}(\omega) \big]=\frac{R}{1+(\omega RC)^2} \ .
\end{equation}
Numerical integration of Eq.~(\ref{gamma.weak}) using Eq.~(\ref{totresR}) is straightforward. Results for $R_K
\Gamma/R$ as a function of $k_B T_\mathrm{env}/\Delta$ are shown in Fig.~\ref{fig:resistives} for various values of the
parameter $\Delta RC/\hbar$. We see that $\Gamma$ increases monotonically with temperature. Also shown is the
asymptotic linear temperature dependence of $\Gamma$ reached for temperatures $k_BT_\textrm{env}\gg\Delta$,
\begin{equation}
\label{gamma.weak.a}\Gamma\simeq2\pi \frac{R}{R_K}\frac{k_BT_{\textrm{env}}}{\Delta} \left[1 - \frac{ \Delta
RC/\hbar}{\sqrt{1 + (\Delta RC/\hbar)^2}}\right].
\end{equation}
As the parameter $\Delta RC/\hbar$ is increased, the slope characterizing the limiting dependence decreases:
photon-assisted inelastic tunneling is effectively reduced by increasing the junction capacitance. Note that in the
limit $\Delta RC/\hbar \ll 1$ the result (\ref{gamma.weak.a}) tends to $\Gamma = 2 \pi (R/R_K)
(k_BT_\textrm{env}/\Delta)$, obtained in Ref.~\onlinecite{main}.

\begin{figure*}[ht]
\centering \subfigure[] {\psfrag{01}[tc][tc][0.85]{$\qquad 0.05$} \psfrag{02}[tc][tc][0.85]{$\qquad 0.1$}
\psfrag{03}[tc][tc][0.85]{$\qquad 0.5$} \psfrag{04}[tc][tc][0.85]{$\qquad 1$} \psfrag{05}[tc][tc][0.85]{$\qquad 5$}
\psfrag{06}[tc][tc][0.85]{$\qquad 10$} \psfrag{Leglab}[tc][tc][0.9]{$E_C/\Delta$}
\psfrag{xxx}[bc][bc][1.3]{$k_BT_\textrm{env}/\Delta$}\psfrag{yyy}[Bc][Bc][1.3]{$\Gamma$}
\includegraphics[trim=-10 0 10 0,height=0.305\textwidth]{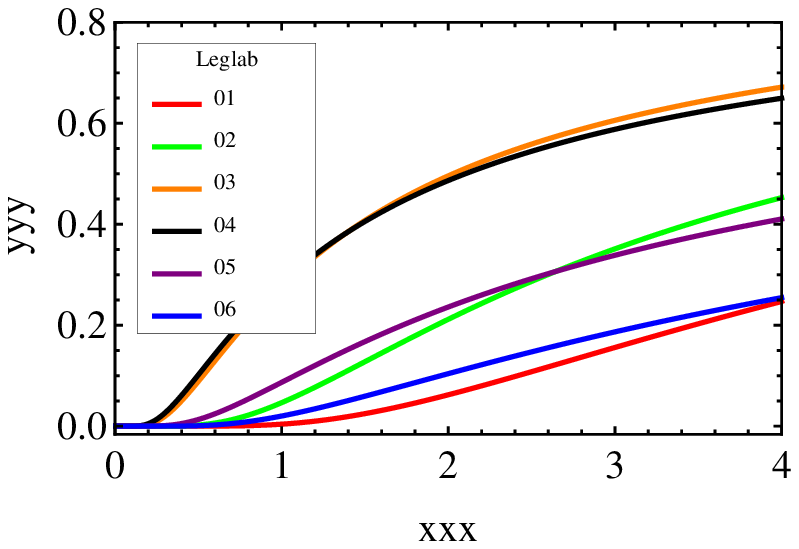}\label{fig:strong1}}
\hspace{-3mm} \subfigure[] {\newcommand{\myfbox}{\fbox} \psfrag{01}[tc][tc][0.85]{$\qquad 0.3$}
\psfrag{02}[tc][tc][0.85]{$\qquad 0.5$} \psfrag{03}[tc][tc][0.85]{$\qquad 1.0$} \psfrag{04}[tc][tc][0.85]{$\qquad 2.0$}
\psfrag{05}[tc][tc][0.85]{$\qquad 3.0$} \psfrag{06}[tc][tc][0.85]{$\qquad
5.0$}\psfrag{Leglab}[tc][tc][0.9]{$k_BT_\textrm{env}/\Delta$}
\psfrag{xxx}[bc][bc][1.3]{$E_C/\Delta$}\psfrag{yyy}[Bc][Bc][1.3]{$\Gamma$}
\includegraphics[trim=-7 0 10 0,height=0.31\textwidth]{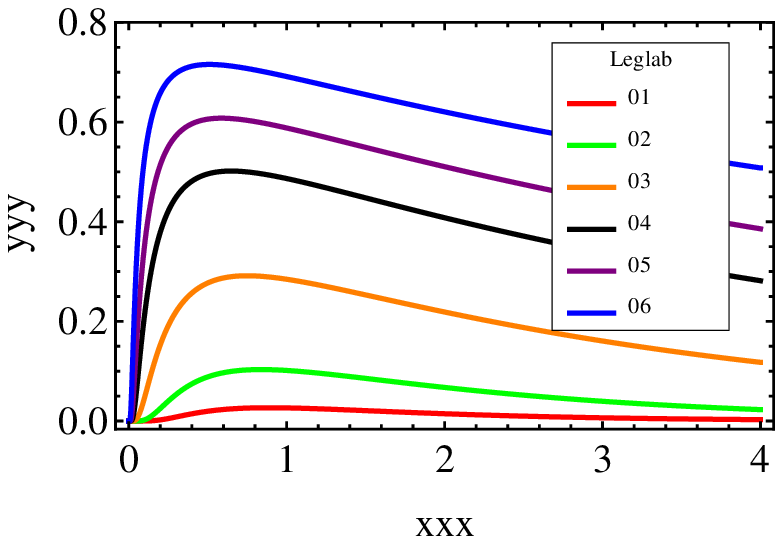}\label{fig:strong2}}
\caption{(a) Plot of the parameter $\Gamma$ as a function of $k_B T_\textrm{env}/\Delta$ obtained considering the
numerical integration of Eq.~(\ref{gamma.strong}). Each curve refers to a certain fixed value of the ratio $E_C/\Delta$
(see legend). (b) Numerical plot of the same quantity, Eq.~(\ref{gamma.strong}), as a function of $E_C/\Delta$ for
different values of the ratio $k_B T_\textrm{env}/\Delta$, as indicated.}
\end{figure*}

\subsection{Subgap Leakage Current: Strong Coupling}\label{Sec.NIS.strong}

We do not aim to present a general analysis in the strong coupling limit. In the particular case where
$\Re\mbox{e}[Z_\textrm{tot}(\omega)]$ is strongly peaked around $\omega=0\,$, the probability density $P(E)$ can be
calculated explicitly \cite{IngNaz} and results for the parameter $\Gamma$ obtained. Let us illustrate this by
considering a purely resistive environment. When the resistance is big, $R\gg R_K \Delta/k_B T_\textrm{env}$ (see
Sec.\ref{Sec.NIS.w&s}), the impedance (\ref{totresR}) becomes
\begin{equation}
\label{strong.cond} \Re\mbox{e}[Z_\textrm{tot}(\omega)]\simeq\Bigg(\frac{\pi}{C}\Bigg) \ \delta(\omega) \ .
\end{equation}
As a result, the function $P(E)$ is given by
\begin{equation}
\label{PEhighImp} P(E)\simeq\frac{1}{\sqrt{4\pi k_BT_{\textrm{env}}E_C}} \ \exp \left[ -\frac{\Big(E-E_C
\Big)^2}{4k_BT_{\textrm{env}}E_C} \ \right] \ .
\end{equation}
Here we defined the charging energy  $E_C = e^2/2C$. Inserting the function (\ref{PEhighImp}) in equation
(\ref{gamma}), we find
$$
\Gamma=\frac{1}{\sqrt{\pi E_C k_BT_{\textrm{env}}}} \int_{\Delta}^{+\infty}dE \ N_S(E) \ \times
$$
\begin{equation}
\label{gamma.strong} \times \exp\left[ -\frac{\Big( E+E_C\Big)^2}{4E_C k_BT_{\textrm{env}}} \right]\ .
\end{equation}
Note that this result depends on $R\,$ implicitly only, through the requirement $R\gg R_K \Delta/k_B T_\textrm{env}\,$.
Direct numerical integration of (\ref{gamma.strong}) yields $\Gamma$ as a function of $k_BT_\mathrm{env}/\Delta$ and
$E_C/\Delta$, as shown in Figs.~\ref{fig:strong1} and \ref{fig:strong2}.  Some remarks are in order at this point.
First of all, for $E_C\ll \Delta$, the integral in Eq.~(\ref{gamma.strong}) can be evaluated approximately, $\Gamma
\simeq e^{-\Delta^2/k_BT_\mathrm{env} E_C}$. As in the weak coupling regime, large values of the capacitance lead to a
reduction of the parameter $\Gamma$. Upon increasing the ratio $E_C/\Delta$, $\Gamma$ will first increase, then it
decreases again when $E_C/\Delta > 1$, which is a manifestation of the Coulomb blockade. As a function of temperature,
$\Gamma$ increases monotonically, similarly to the weak coupling limit. However, rather than reaching an asymptotic
linear dependence, $\Gamma$ saturates at $\Gamma = 1$ for temperatures $k_BT_\mathrm{env} E_C \gg \Delta^2$: the noise
is so strong that features of the order of the gap $\Delta$ are washed out.

\section{NIS Junction coupled to a high-temperature environment by means of a transmission line}

In the previous section we have studied the subgap leakage current in a NIS junction which is directly coupled to the
external environment $Z(\omega)$. We have seen that a reduction of the subgap leakage current is possible when the
capacitance of the junction, $C$, is increased and/or the resistance of the environment, $R$, is decreased.
Unfortunately, in real experiments $R$, and in particular $C$, cannot be chosen arbitrarily and one needs other means
to achieve the accuracy requirements for the aforementioned NIS junction's applications. We therefore consider the
circuit of Fig.~\ref{fig2} where the junction is indirectly coupled to the external environment via a cold, lossy
transmission line acting as a filter.

\subsection{Voltage Fluctuations in the Presence of a Transmission Line}\label{Sec.fluc.trans}

\begin{figure}[t]
\centering
\begin{psfrags}\psfrag{dV}[tc][tc][1.3]{$\delta V$}
\psfrag{dVj}[tc][tc][1.3]{$\delta V_J$}
   \psfrag{dI}[tc][tc][1.3]{$\delta I$}
   \psfrag{Zj}[tc][tc][1.3]{$Z_J(\omega)$}
   \psfrag{par}[tc][tc][1.3]{$R_0\, , C_0\, , L_0$}
   \psfrag{Te}[tc][tc][1.3]{$T_\textrm{env}$}
   \psfrag{Z}[tc][tc][1.3]{$Z(\omega)$}
   \psfrag{x}[tc][tc][1.2]{$x$}
   \psfrag{ll}[tc][tc][1.2]{$\ell$}
   \psfrag{0}[tc][tc][1.2]{$0$}
\includegraphics[height=3.25cm]{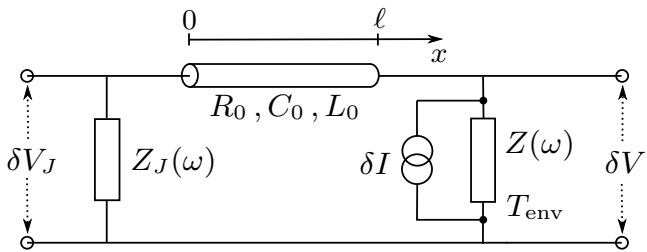}
\caption{{\footnotesize Sketch of the circuit discussed in Sec.~\ref{Sec.fluc.trans}.}}\label{fig:trans}\end{psfrags}
\end{figure}

In order to find the correlation function $J(t)$ in the presence of the transmission line, we solve the intermediate
problem of the propagation of the noise generated by the high-temperature environment with impedance $Z(\omega)$
through the line towards the junction, as shown in Fig.~\ref{fig:trans}. The line has a length length $\ell$ and is
described by the parameters $R_0$, $C_0$ and $L_0$, the resistance, the capacitance and the inductance per unit length
respectively. We ignore the thermal noise produced by the impedance $Z_J(\omega)$ and by the line, assuming both
components at zero temperature. The high-temperature element produces current noise $\delta I$ which in turn induces
voltage noise $\delta V$.

To understand how the potential drop $\delta V_J$ across $Z_J(\omega)$ is connected to $\delta V=Z(\omega)\,\delta I$,
we start considering the potential $V(x)$ and the current $I(x)$ at a given point $x$ along the transmission line. They
satisfy the two partial differential equations,
\begin{align*}
&\frac{\partial V(x)}{\partial x}=-I(x)\Big[R_0-i  \omega  L_0 \Big]\, ,\\
&\frac{\partial I(x)}{\partial x}=i  \omega   C_0 \  V(x) \ .
\end{align*}
Combining them one obtains the wave equation
\begin{equation}\label{transline.eq.wave}
\frac{\partial^2 V(x)}{\partial x^2}=-K^2(\omega) \ V(x)\, ,
\end{equation}
where $K^2(\omega)=\omega^2L_0C_0+i\omega R_0C_0$ is the wave vector squared of the
signal which propagates along the line. A general solution of
Eq.~(\ref{transline.eq.wave}) is given by
\begin{equation}\label{volt.wave}
V(x)=A \ e^{iK(\omega)x}+B \ e^{-iK(\omega)x} \ .
\end{equation}
Consequently the current along the line is
\begin{equation}\label{curr.wave}
I(x)=\frac{1}{Z_\infty(\omega)}\Big[A e^{iK(\omega)x}-B e^{-iK(\omega)x}\Big]\, ,
\end{equation}
with $Z_\infty(\omega)=i\big(R_0-i\omega L_0 \big)/K(\omega)$. The parameters $A$ and $B$ can be determined by means of
the boundary conditions
\begin{align*}
&V(\ell)=Z(\omega) \Big[ I(\ell)+\delta I \Big]=Z(\omega) \ I(\ell)+\delta V\\
&V(0)=-Z_J(\omega) \ I(0) \ .
\end{align*}
As a result, the potential drop $\delta V_J=V(0)=A+B$ across the impedance $Z_J(\omega)$ depends on the noise $\delta
V$ according to the relation
\begin{equation}\label{transline.pot}
\delta V_J= T(\omega) \ \delta V \ .
\end{equation}
In this last equation we introduced $T(\omega)\,$, the transmission function
$$
T(\omega)=\frac{2 \ Z_\infty(\omega) \
Z_J(\omega)}{\Big[Z_\infty(\omega)+Z(\omega)\Big]\Big[Z_\infty(\omega)+Z_J(\omega)\Big]} \ \times
$$
\begin{equation}\label{transline.param}
\times \ \frac{1}{e^{-iK(\omega)\ell}-\lambda_1(\omega) \ \lambda_2(\omega) \ e^{iK(\omega)\ell}}\, ,
\end{equation}
where
$$
\lambda_1(\omega)=\frac{Z_\infty(\omega)-Z(\omega)}{Z_\infty(\omega)+Z(\omega)}\qquad
\lambda_2(\omega)=\frac{Z_\infty(\omega)-Z_J(\omega)}{Z_\infty(\omega)+Z_J(\omega)}
$$
are the reflection coefficients. Assuming that the potential $\delta V$ satisfies the quantum fluctuation-dissipation
theorem,
$$
\Big\langle \delta V(t) \ \delta V(0) \Big\rangle_\omega=2\hbar\omega \
\frac{\Re\mbox{e}\big[Z(\omega)\big]}{1-e^{-\beta_\textrm{env}\hbar\omega}} \ ,
$$
the spectral density function of the potential (\ref{transline.pot}) is
\begin{equation}\label{transline.spectr}
\Big\langle \delta V_J(t) \ \delta V_J(0)\Big\rangle_\omega=\Big| T(\omega) \Big|^2\ 2\hbar\omega \
\frac{\Re\mbox{e}\big[Z(\omega)\big]}{1-e^{-\beta_\textrm{env}\hbar\omega}} \ .
\end{equation}
This expression describes the propagation of the noise from $Z(\omega)$ to the noiseless impedance $Z_J(\omega)$
through the noiseless transmission line.

\subsection{Correlation Function for the Transmission Line Circuit}

We use Eq.~(\ref{transline.spectr}) to calculate the modified correlation function $J_T(t)$ which appears in
Eq.~(\ref{PE}). According to Ref.~\onlinecite{IngNaz}, $J(t)$ is defined as the correlation function
\begin{equation}\label{J.corr}
J(t)\equiv\Big\langle\varphi_J(t)\varphi_J(0)-\varphi_J(0)\varphi_J(0)\Big\rangle \ ,
\end{equation}
where the phase $\varphi_J(t)$ is the time integral of the potential $\delta V_J(t)$
across the NIS junction,
$$
\varphi_J(t)\equiv \ \frac{e}{\hbar} \int_{-\infty}^t \delta V_J(\tau) \ d\tau\, .
$$
In other words
\begin{equation}\label{phi.Vu}
\Big\langle \varphi_J(t) \ \varphi_J(0)\Big\rangle_\omega= \ \Big(\frac{e}{\hbar} \Big)^2\frac{1}{\omega^2} \
\Big\langle \delta V_J(t) \ \delta V_J(0)\Big\rangle_\omega\, .
\end{equation}
Using the fluctuation-dissipation relation (\ref{transline.spectr}) in
 (\ref{phi.Vu}), we rewrite Eq.~(\ref{J.corr}) as a function of
$T(\omega)$, $Z(\omega)$ and $T_\textrm{env}$. Taking the impedance $Z_J(\omega)$ to be
the one of a capacitance $C$, the modified function $J_T(t)$ reads
$$
J_T(t)= \ 2 \ \int_0^{+\infty}\frac{d\omega}{\omega} \ \Big| T_C(\omega) \Big|^2 \
\frac{\Re\mbox{e}\big[Z(\omega)\big]}{R_K} \ \times
$$
\begin{equation}\label{J(t)2temp}
\times\Bigg\{\coth\Big(\frac{1}{2}\beta_\textrm{env}\hbar\omega\Big)\Big[\cos\big(\omega t\big)-1\Big]-
i\sin\big(\omega t \big)\Bigg\} \ .
\end{equation}
Here $T_C(\omega)$ is the function $T(\omega)$, Eq.~(\ref{transline.param}), with $Z_J(\omega)=Z_C(\omega)=-1/i\omega
C$. Since the transmission line is considered noiseless, its temperature $T_\textrm{line}$ should be low,
$T_\textrm{line}\ll\Delta/k_B$. In what follows we set $T_\textrm{line}=T_\textrm{junc}=0$.

\subsection{The Transmission Function}\label{Sec.trans.function}

In order to understand the effect of the insertion of the transmission line in the circuit of Fig.~\ref{fig1}, a
discussion about the general behavior of $T_C(\omega)$ is necessary. In general, the modulus squared of the
transmission function (\ref{transline.param}) is characterized by a series of resonance peaks, whose properties depend
on $\ell,R_0,C_0$, and $L_0$ as well as on the external impedance $Z\big(\omega\big)$. To have an idea of the behavior
of $\big| T_C(\omega) \big|^2$, let us consider the case of a purely resistive environment, $Z\big(\omega\big)=R\,$.

\begin{figure*}[!t]
\begin{psfrags}
\subfigure[]{\psfrag{z07}[tc][tc][0.85]{$\qquad z_0=7$} \psfrag{z05}[tc][tc][0.85]{$\qquad z_0=5$}
\psfrag{z04}[tc][tc][0.85]{$\qquad z_0=4$} \psfrag{z03}[tc][tc][0.85]{$\qquad z_0=3$}
\psfrag{Lorentzian}[tc][tc][0.85]{$\phantom{zz}$Lorentzian}\psfrag{xxx}[bc][bc][1.4]{$\omega RC$}
\psfrag{yyy}[Bc][tc][1.4]{$\big|T_C(\omega)\big|^2$}
\includegraphics[trim=-10 0 10 0,width=0.48\textwidth]{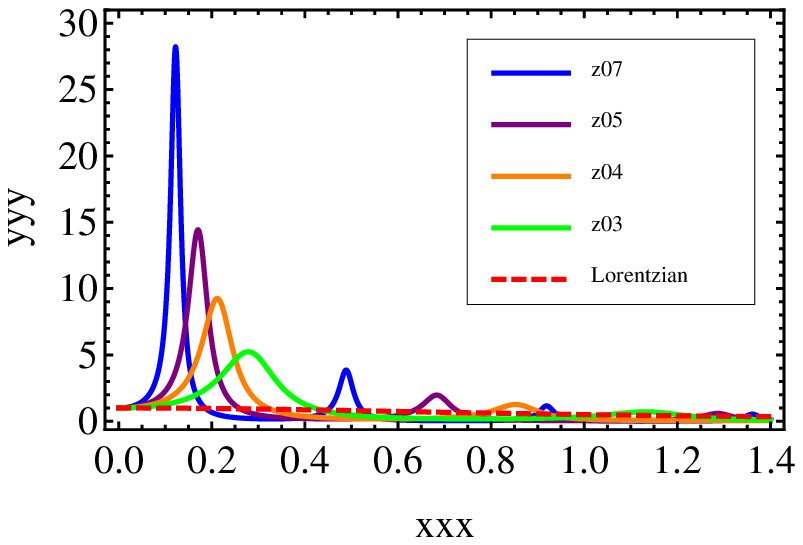}\label{fig:plot1}}\subfigure[]{\psfrag{c010}[tc][tc][0.85]{$\qquad
c_0=10$} \psfrag{c07}[tc][tc][0.85]{$\qquad c_0=7$} \psfrag{c05}[tc][tc][0.85]{$\qquad c_0=5$}
\psfrag{c03}[tc][tc][0.85]{$\qquad c_0=3$}
\psfrag{Lorentzian}[tc][tc][0.85]{$\phantom{zz}$Lorentzian}\psfrag{xxx}[bc][bc][1.4]{$\omega RC$}
\psfrag{yyy}[Bc][tc][1.4]{$\big|T_C(\omega)\big|^2$}
\includegraphics[trim=-10 0 10 0,width=0.48\textwidth]{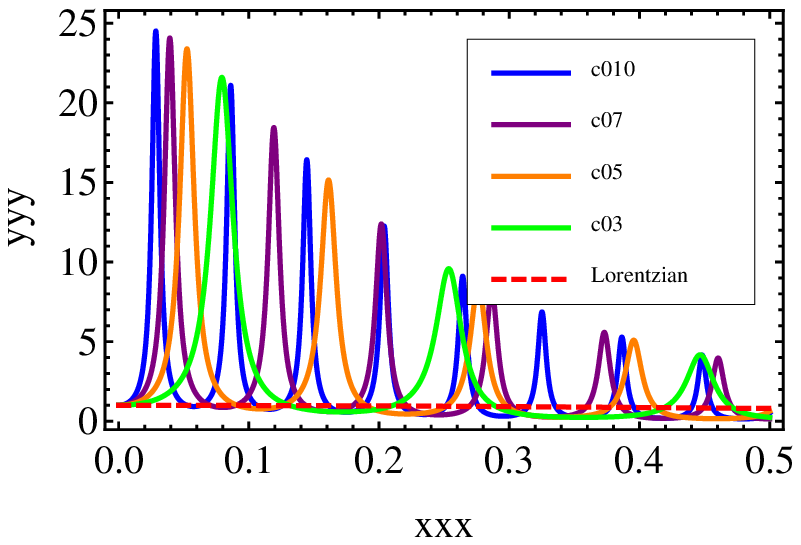}\label{fig:plot2}}\\\subfigure[]{\psfrag{z008}[tc][tc][0.85]{$\qquad z_0=0.8$}
\psfrag{z006}[tc][tc][0.85]{$\qquad z_0=0.6$} \psfrag{z005}[tc][tc][0.85]{$\qquad z_0=0.5$}
\psfrag{z003}[tc][tc][0.85]{$\qquad z_0=0.3$}
\psfrag{Lorentzian}[tc][tc][0.85]{$\phantom{zz}$Lorentzian}\psfrag{xxx}[bc][bc][1.4]{$\omega RC$}
\psfrag{yyy}[Bc][tc][1.4]{$\big|T_C(\omega)\big|^2$}
\includegraphics[trim=-8 0 8 0,width=0.48\textwidth]{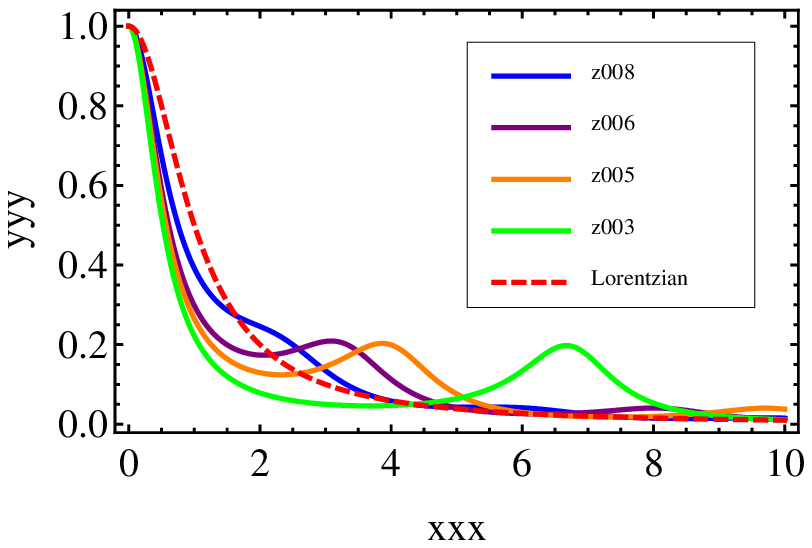}\label{fig:plot3}}\subfigure[]{\psfrag{c010}[tc][tc][0.85]{$\qquad c_0=10$}
\psfrag{c07}[tc][tc][0.85]{$\qquad c_0=7$} \psfrag{c05}[tc][tc][0.85]{$\qquad c_0=5$}
\psfrag{c03}[tc][tc][0.85]{$\qquad c_0=3$}
\psfrag{Lorentzian}[tc][tc][0.85]{$\phantom{zz}$Lorentzian}\psfrag{xxx}[bc][bc][1.4]{$\omega RC$}
\psfrag{yyy}[Bc][tc][1.4]{$\big|T_C(\omega)\big|^2$}
\includegraphics[trim=-10 0 10 0,width=0.48\textwidth]{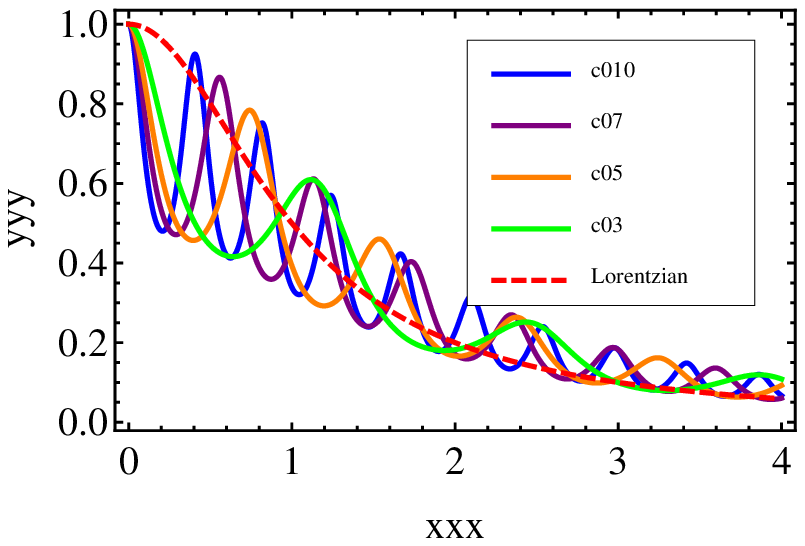}\label{fig:plot4}}\\\subfigure[]{\psfrag{r01}[tc][tc][0.85]{$\qquad r_0=1$}
\psfrag{r02}[tc][tc][0.85]{$\qquad r_0=2$} \psfrag{r03}[tc][tc][0.85]{$\qquad r_0=3$}
\psfrag{r04}[tc][tc][0.85]{$\qquad r_0=4$} \psfrag{Lorentzian}[tc][tc][0.85]{$\phantom{zz}$Lorentzian}
\psfrag{xxx}[bc][bc][1.4]{$\omega RC$} \psfrag{yyy}[Bc][tc][1.4]{$\big|T_C(\omega)\big|^2$}
\includegraphics[trim=-10 0 10 0,width=0.48\textwidth]{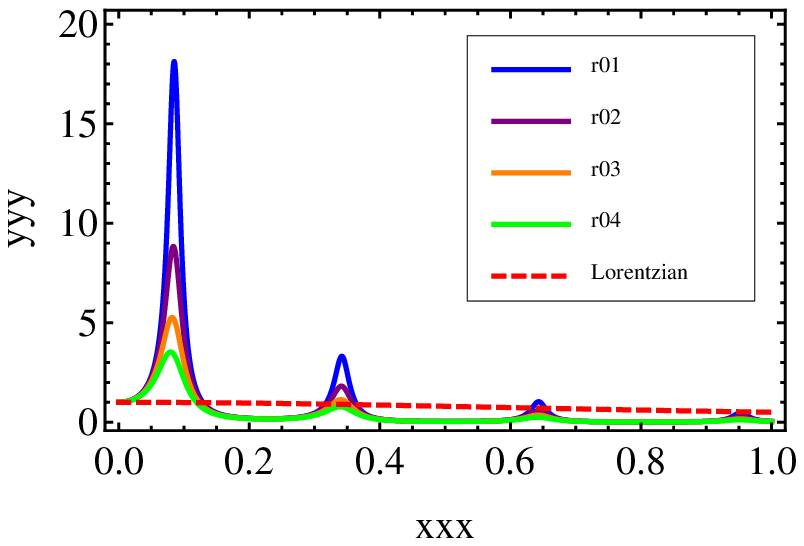}\label{fig:plot5}}\subfigure[]{\psfrag{r07}[tc][tc][0.85]{$\qquad r_0=7$}
\psfrag{r09}[tc][tc][0.85]{$\qquad r_0=9$} \psfrag{r010}[tc][tc][0.85]{$\qquad r_0=10$}
\psfrag{r012}[tc][tc][0.85]{$\qquad r_0=12$} \psfrag{Lorentzian}[tc][tc][0.85]{$\phantom{zz}$Lorentzian}
\psfrag{xxx}[bc][bc][1.4]{$\omega RC$} \psfrag{yyy}[Bc][tc][1.4]{$\big|T_C(\omega)\big|^2$}
\includegraphics[trim=-10 0 10 0,width=0.48\textwidth]{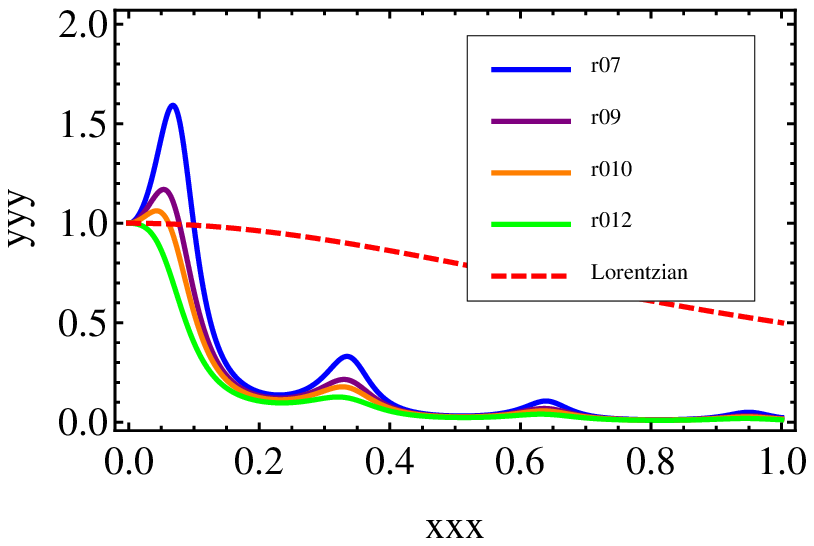}\label{fig:plot7}}
\end{psfrags}
\caption{\label{fig:multi_panel}Plots of the transmission function $\big| T_C(\omega) \big|^2$ as a function of the
dimensionless variable $\omega RC$. Each panel corresponds to a different set of the parameters $z_0$, $c_0$ and $r_0$:
(a) $r_0=0$ , $c_0=1$ , $z_0=(7, 5, 4, 3)$; (b) $r_0=0$ , $z_0=5$ , $c_0=(10, 7, 5, 3)$; (c) $r_0=0$ , $c_0=1$ ,
$z_0=(0.8, 0.6 ,0.5 ,0.3)$; (d) $r_0=0$ , $z_0=0.7$ , $c_0=(10, 7, 5, 3)$; (e) $z_0=10$ , $c_0=1$ , $r_0=(1, 2, 3, 4)$;
(f) $z_0=10$ , $c_0=1$ , $r_0=(7, 9, 10, 12)$.}
\end{figure*}

Figure \ref{fig:multi_panel} illustrates the behavior of $\big| T_C(\omega) \big|^2$ as a function of $\omega RC$ for
different values of the dimensionless parameters $z_0=\sqrt{L_0/C_0}/R\,$, $c_0=\ell C_0/C$ and $r_0=\ell R_0/R\,$.
Also shown is the Lorentzian result
\begin{equation}
\label{trans.sostitution} \big| T_C(\omega) \big|^2 = 1/[1+ (\omega RC)^2]
\end{equation}
found for $\ell =0$, {\em i.e.}, in the absence of the transmission line. In order for the line to be an efficient
filter, we require $\big| T_C(\omega) \big|^2$ to be below this Lorentzian curve in the relevant frequency ranges. We
see that both the position and the width of the resonance peaks are proportional to $\pi/2 c_0 z_0\,$: the longer is
the transmission line, the denser around zero and the sharper are the peaks. Their height decreases rapidly as the
dimensionless frequency $\omega RC$ is increased. This can be seen in particular when the line has no losses, $r_0 =
0$, see Figs.~\ref{fig:plot1} -- \ref{fig:plot4}. Although the Lorentzian curve is approached for lossless lines when
$c_0$ or $z_0$ is reduced, we observe no real reduction below it.

\begin{figure}[!t]
\begin{psfrags}
\psfrag{r01}[tc][tc][0.85]{$\qquad r_0=1$} \psfrag{r02}[tc][tc][0.85]{$\qquad r_0=2$}
\psfrag{r03}[tc][tc][0.85]{$\qquad r_0=3$} \psfrag{r04}[tc][tc][0.85]{$\qquad r_0=4$}
\psfrag{r050}[tc][tc][0.85]{$\qquad r_0=50$}
\psfrag{Lorentzian}[tc][tc][0.85]{$\phantom{zz}$Lorentzian}\psfrag{xxx}[bc][bc][1.4]{$\omega RC$}
\psfrag{yyy}[Bc][tc][1.4]{$\big|T_C(\omega)\big|^2$}
\includegraphics[trim=-10 5 10 2,height=0.29\textwidth]{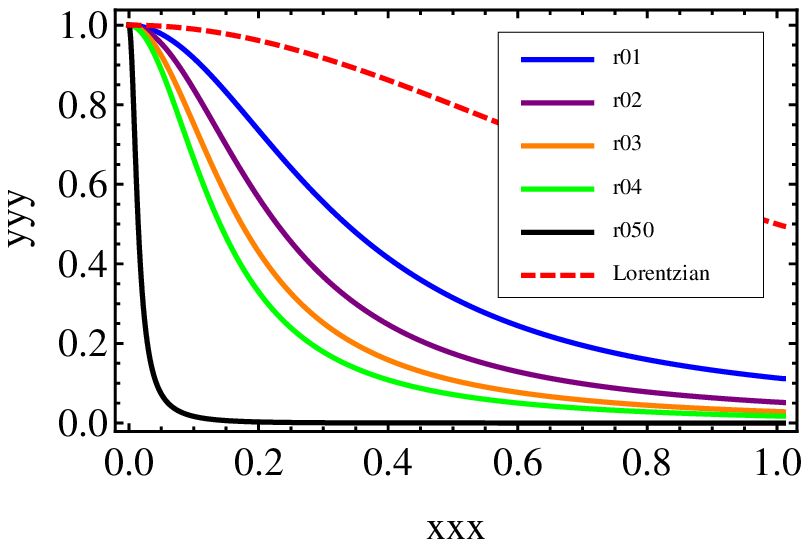}
\end{psfrags}
\caption{Plot of the transmission function $\big| T_C(\omega) \big|^2$ as a function of the dimensionless variable
$\omega RC$ for different values of the parameter $r_0$. The other parameters are: $c_0=1\,$, $z_0=0.7\,$. Also shown
is the Lorentzian corresponding to the function $\big| T_C(\omega) \big|^2$ in the limit $\ell\rightarrow 0\,$, given
by Eq.~(\ref{trans.sostitution}).}\label{fig:transfunc}
\end{figure}

A significant reduction of the height of the peaks is possible if the line which connects
the NIS junction and the environment is lossy, $r_0>0$. Indeed, we see from
Figs.~\ref{fig:plot5} and \ref{fig:plot7} that the bigger is $r_0$ the smaller are the
local maxima of $\big| T_C(\omega) \big|^2\,$. Moreover, the transmission function is
even much smaller than $1/\big[1+\big(\omega RC\big)^2\big]\,$ when the condition $r_0\gg
z_0$ is satisfied, as is seen in Figs.~\ref{fig:plot7} and \ref{fig:transfunc}.
Therefore, within this particular limit, the insertion of a resistive transmission line
may be convenient.

\subsection{Subgap Leakage Current: Weak Coupling}

\begin{figure}[!t]
\newcommand{\myfbox}{\fbox}
\psfrag{01}[tc][tc][0.85]{$\qquad\qquad \ell=0$} \psfrag{02}[tc][tc][0.85]{$\qquad\qquad 10^6\unit{\ohm/m}$}
\psfrag{03}[tc][tc][0.85]{$\qquad\qquad 10^9\unit{\ohm/m}$} \psfrag{04}[tc][tc][0.85]{$\qquad\qquad
10^{10}\unit{\ohm/m}$} \psfrag{05}[tc][tc][0.85]{$\qquad\qquad 10^{11}\unit{\ohm/m}$}
\psfrag{06}[tc][tc][0.85]{$\qquad\qquad 10^{12}\unit{\ohm/m}$}
\psfrag{xxx}[bc][bc][1.7]{$\rho$}\psfrag{yyy}[Bc][Bc][1.7]{$\tau_\textrm{10\%}(\rho)$}
\includegraphics[trim=0 5 10 4,height=0.29\textwidth]{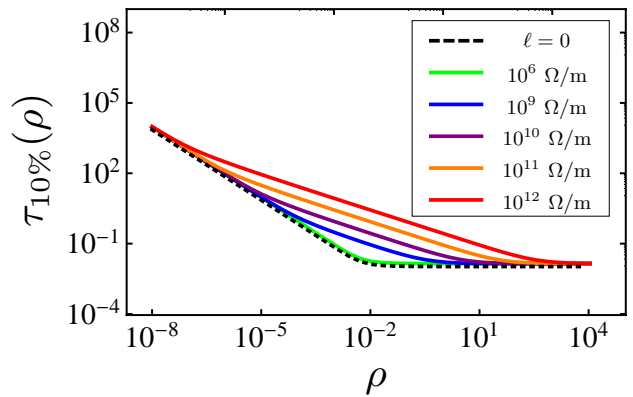}
\caption{Plot of the dimensionless time $\tau_\textrm{\tiny{10\%}}(\rho)$ as a function of the dimensionless resistance
$\rho=R/R_K\,$ for different values of the resistance per unit length, $R_0\,$, of the transmission line (see legend).
Also shown is the curve $\tau_\textrm{\tiny{10\%}}(\rho)\,$, valid for the circuit of Fig.~\ref{fig1} (black dashed
line). The value of the other parameters are: $\Delta\simeq 200\unit{\mu\/eV}$ (energy gap of Aluminum),
$T_\textrm{env}=5\unit{K}\,$, $C=10\unit{f\/F}\,$, $C_0= 6\,\varepsilon_0\,$, $L_0=\mu_0\,$, $\ell=10\unit{\mu m}$.
}\label{fig:resistivesind}
\end{figure}

We expect that the single- and multi-photon regimes, weak and strong coupling
respectively, are strongly related to the resistance per unit length, $R_0\,$. Let us
analyze the situation proceeding as in Sec.\ref{Sec.NIS.w&s}. We consider the function
$\tau_\textrm{\tiny{10\%}}(\rho)$ for a purely resistive environment. In
Fig.~\ref{fig:resistivesind} we plot $\tau_\textrm{\tiny{10\%}}(\rho)$ as a function of
the dimensionless resistance $\rho$ for different values of $R_0\,$. We see that the
lossier the transmission line is, the more the weak coupling region spreads out. The
resistance $\rho_\Delta\,$, given by the intersection between
$\tau_\textrm{\tiny{10\%}}(\rho)$ and the line corresponding to the dimensionless time
$\tau_{\scriptscriptstyle\Delta}=\hbar/\Delta R_K C\,$, significantly shifts towards
higher values of $\rho$ as $R_0$ is increased; the lossy line indeed protects the
junction from the high-temperature external environment. Hereafter, we will therefore
focus on a highly resistive transmission line and only the weak coupling regime will be
treated.

With the help of Eq.~(\ref{J(t)2temp}), the function $P(E)$ for the circuit of
Fig.~\ref{fig2} can be obtained in the weak coupling regime. Proceeding as in
Sec.~\ref{Sec.NIS.weak}, we find
\begin{equation}\label{PEweak2}
P(E)\simeq \ 2 \ \Big| T_C(E/\hbar) \Big|^2 \ \frac{\Re\mbox{e}\big[Z\big(E/\hbar\big)
\big]}{R_K}\Bigg(\frac{1+n\big(E\big)}{E} \Bigg)\, .
\end{equation}
Evaluating the relation (\ref{PEweak2}) for negative energies and inserting the result
into Eq.~(\ref{gamma}), the parameter $\Gamma$ can be written as
\begin{equation}\label{gamma.transm}
\Gamma= 4\int_\Delta^{+\infty} dE \ N_S(E) \ \Big| T_C(E/\hbar) \Big|^2
\frac{\Re\mbox{e}\big[ Z\big(E/\hbar\big)\big]}{R_K} \ \frac{n\big(E\big)}{E} \ .
\end{equation}
We next specialize to the case of large resistance per unit length, $R_0\,$. In order to
obtain a limiting expression for  $\big| T_C(\omega) \big|^2$ for
$R_0\rightarrow\infty\,$, let us assume that the inductive properties of the line are
negligible compared to $R_0\,$. Since the relevant frequency scale is given by
$\Delta/\hbar\,$, this means that the condition $R_0\gg L_0\Delta/\hbar$ should hold.
Within this $RC$ limit, we find that the wave vector $K(\omega)$ of the signal
propagating through the transmission line has an imaginary part equal to $\sqrt{\omega
R_0C_0/2}\,$. As a result, the amplitude of the noise is exponentially attenuated along
the line (see Eqs.~(\ref{volt.wave}) and (\ref{curr.wave})) being proportional to
$\exp{\big[-\ell\sqrt{2\omega R_0 C_0}\big]}$. We see that the bigger $\ell$ and $R_0$
are, the smaller is the voltage noise which reaches the junction. In particular, an
exponential suppression of the propagating signal is achieved when the inequality $\ell\,
\sqrt{2\Delta R_0 C_0/\hbar} \gg1\,$ is valid as well. This additional condition allows
us to write the equation
$$
\Big| e^{-iK(\omega)\ell}-\lambda_1(\omega) \ \lambda_2(\omega) \ e^{iK(\omega)\ell}\Big|^2\simeq \ 4 \ e^{\ell
\sqrt{2\omega R_0 C_0}} \ .
$$
Then the modulus squared of the transmission function $T_C(\omega)$ becomes
\begin{equation}
\label{Tc.approx} \Big|T_C(\omega)\Big|^2\simeq\left|\frac{Z_\infty(\omega) \ Z_C(\omega) \ \ e^{-\ell \sqrt{2\omega
R_0 C_0}/2}}{\Big[Z_\infty(\omega)+Z(\omega)\Big]\Big[Z_\infty(\omega)+Z_C(\omega)\Big]}\right|^2
\end{equation}
where $Z_\infty(\omega)\simeq(1+i)\sqrt{R_0/2\omega C_0}$ for a line in the $RC$ limit.
Combining the two conditions used so far, we find that the approximated function
(\ref{Tc.approx}) holds when the resistance of the transmission line, $\ell R_0\,$, is
much bigger than its characteristic impedance $Z_\infty=\sqrt{L_0/C_0}\,$.

\begin{figure}[!t]
\psfrag{first}[tc][tc][1]{\setlength\fboxsep{1.5pt}{\colorbox{white}{\framebox{$10^{11}$}}}}
\psfrag{second}[tc][tc][1]{\setlength\fboxsep{1.5pt}{\colorbox{white}{\framebox{$10^{10}$}}}}
\psfrag{third}[tc][tc][1]{\setlength\fboxsep{1.5pt}{\colorbox{white}{\framebox{$10^9$}}}}
\psfrag{forth}[tc][tc][1]{\setlength\fboxsep{1.5pt}{\colorbox{white}{\framebox{$10^8$}}}}
\psfrag{fifth}[tc][tc][1]{\setlength\fboxsep{1.5pt}{\colorbox{white}{\framebox{$10^7$}}}}
\psfrag{sixth}[tc][tc][1]{\setlength\fboxsep{1.5pt}{\colorbox{white}{\framebox{$10^6$}}}}
\psfrag{seventh}[tc][tc][1]{\setlength\fboxsep{1.5pt}{\colorbox{white}{\framebox{$10^{12}$}}}}
\psfrag{xxx}[bc][bc][1.4]{$\ell\unit{(m)}$}\psfrag{yyy}[Bc][Bc][1.4]{$\Gamma$}
\includegraphics[trim=12 5 10 4,height=0.3\textwidth]{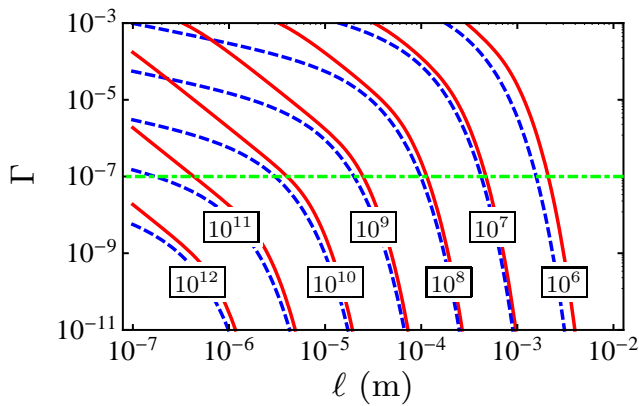}
\caption{Plot of the parameter $\Gamma$, Eq.~(\ref{gamma.transm}), as a function of the length of the transmission line
$\ell$. The red solid line is obtained by means of the numerical integration of Eq.~(\ref{gamma.transm}) for a purely
resistive environment. The blue dashed line is the plot of the asymptotic parameter given by the
Eq.~(\ref{gamma.decay}). These two curves are plotted for different values of the resistance per unit length
$R_0\unit{(\ohm/ m)}$ (as indicated in the graph). All the plots are obtained considering the gap parameter of the
aluminum, $\Delta\simeq 200\unit{\mu\/eV}$. The other parameters are: $T_\textrm{env}=5\unit{K}$, $C=10\unit{f\/F}$,
$R=10\unit{\ohm}$ , $C_0= 6\,\varepsilon_0$, $L_0=\mu_0$.}\label{fig.transline}
\end{figure}

Increasing the resistance per unit length, $R_0\,$, one also expects that interference
effects become negligible. Indeed, when $R_0$ is very big, the amplitude of the signal
across the junction is much smaller than its starting value and its reflected counterpart
vanishes rapidly before reaching the noise source again. In terms of our description of
the transmission line given in Sec.~\ref{Sec.fluc.trans}, this happens when the
reflection coefficients $\lambda_1(\omega)$ and $\lambda_2(\omega)$ tend to 1. In fact,
in this limit, the potential drop (\ref{volt.wave}) tends to 0 across the junction and to
$\delta V$ across the impedance $Z(\omega)\,$. For a purely resistive environment, this
 regime is reached when $R_0$ is such that the two inequalities $ R^2 \ll
\hbar R_0/2 C_0\Delta$ and $\Delta R_0 C^2/\hbar C_0 \gg 2\,$ hold, in other words, when the resistance of the
environment, $R\,$, is much smaller than $R_0C/2C_0\,$. Equation~(\ref{Tc.approx}) reduces to the asymptotic expression
$$
\Big|T_C(\omega)\Big|^2\simeq\frac{e^{-\ell\sqrt{2 \omega R_0 C_0}}}{1+\omega R_0 C^2/C_0}  \ .
$$
By means of this last relation, the integral in Eq.~(\ref{gamma.transm}) can be evaluated
approximately with the result
$$
\Gamma\simeq \  \ 4 \ \frac{R}{R_K} \ \frac{1}{e^{\Delta/k_BT_\textrm{env}}-1} \ \times
$$
\begin{equation}\label{gamma.decay}
\times \ \sqrt{\frac{\pi}{\ell\sqrt{2 \Delta R_0 C_0/\hbar}}} \ \ \frac{e^{-\ell\sqrt{2 \Delta R_0
C_0/\hbar}}}{1+\Delta R_0 C^2/\hbar C_0}\  \ .
\end{equation}
We see that this asymptotic parameter $\Gamma$ decreases exponentially in terms of $\ell$ and $R_0\,$; the dependence
on the junction capacitance $C$ is rather weak. The insertion of a highly resistive and noiseless transmission line
between the NIS junction and the high-temperature environment indeed helps to suppress the subgap leakage current. The
plot of Fig.~\ref{fig.transline} shows the exponential decay for a set of values of $R_0$ and $\ell$ that can be used
in real experiments. Particularly interesting is the region where $10^8\unit{\ohm /m}\lesssim R_0 \lesssim
10^{10}\unit{\ohm /m}$ and $10\unit{\mu m}\lesssim \ell \lesssim 10^2 \unit{\mu m}\,$. A transmission line whith these
values of $R_0$ and $\ell$ allows one to go far below $\Gamma\simeq\gamma_\textrm{\tiny{Dynes}}\sim 10^{-7}\,$,
\emph{i.e.}, a value of $\Gamma$ which guarantees the achievement of the accuracy requirements for the superconducting
gap-based technological applications of the NIS junction~\cite{SINIS2}.

\section{Multi-particle tunneling}\label{Sec.multi}

Our analysis focuses on the single-particle subgap current through the NIS junction. We
ignore the contribution due to higher order processes in tunneling, such as Andreev
reflection~\cite{andreev1,andreev2,andreev3}. Hence, in order to establish the validity
of our single-particle tunneling assumption, one has to compare the parameter $\Gamma$
characterizing the leakage current with the dimensionless Andreev subgap conductance $g_A
= G_A R_T$. In ballistic junctions, second-order perturbation theory yields the standard
two-particle subgap conductance
\begin{equation}\label{andreev.curr1}
G_A \simeq R_K/[R_T^2 (k_F^2 S)],
\end{equation}
where $k_F^2 S$ is the number of conduction channels in the tunnel barrier. Two-electron
tunneling can be ignored as long as $\Gamma > R_K/R_T k_F^2 S$. Typical
estimates~\cite{andreev3} yield $R_K/R_T k_F^2 S \sim 10 ^{-7}$.

On the other hand, in the diffusive case the electrons reflected by the barrier are
backscattered by the impurities randomly situated close to the barrier in the normal
metal. Interference between the electrons in a region characterized by the coherence
length $\xi_N=\sqrt{\hbar D/\max{\{eV,k_BT_\textrm{jun} \}}}\,$, where $D$ is the
diffusion coefficient, affects the two-particle tunneling probability\cite{andreev4}. As
a result, $G_A$ is given by
\begin{equation}\label{andreev.curr2}
G_A \simeq R_N/R_T^2
\end{equation}
where $R_N$ is the resistance of the diffusive normal metal over a length $\xi_N$.
General estimates are hard to give in this situation, since the result is strongly
geometry-dependent; the condition $\Gamma
> R_N/R_T$ will be more stringent than the one for the ballistic case, especially
under subgap conditions where $\xi_N$ and hence $R_N$ can be large.

Should Andreev reflection become dominant, one can always suppress it efficiently using the Coulomb blockade
feature\cite{andreev3} that suppresses two-particle tunneling more strongly than single-particle tunneling.

\section{Conclusions}

In conclusion, we studied the single-particle tunneling current through a voltage-biased NIS junction. Due to the
presence of the superconducting energy gap $\Delta$ in the BCS density of states, when the junction is kept at the
temperature $T_\textrm{jun}\ll\Delta/k_B$ no current is expected to flow within the subgap region
$-\Delta<eV<\Delta\,$. Actually, even if the higher order tunneling processes are suppressed, a small subgap current is
still measured experimentally. This leakage current limits the accuracy in applications involving NIS junctions. The
origin of the leakage current is the exchange of energy exceeding the gap $\Delta$ between the junction and the
external high-temperature environment in which it is embedded. We studied this mechanism analytically and numerically.
In particular, we found that a cold and lossy transmission line inserted between the junction and the environment
reduces exponentially the subgap leakage current acting as a filter. This indirect configuration helps to achieve the
required suppression of noise.

{\bf Acknowledgments.} The authors thank W. Belzig and G. Rastelli for useful discussions. Financial support from the
Marie Curie Initial Training Network (ITN) Q-NET (project no. 264034), from Institut Universitaire de France and from
Aalto University's ASCI visiting professor programme is gratefully acknowledged. The work was partially supported by
the Academy of Finland through its LTQ (project no. 250280) COE grant (VFM and JPP), and the National Doctoral
Programme in Nanoscience, NGS-NANO (VFM).

\end{document}